\documentclass[aps,pra,twocolumn,showpacs,groupedaddress]{revtex4}
\usepackage[colorlinks=true, linkcolor=red, citecolor=blue, urlcolor=blue]{hyperref}
\usepackage[utf8]{inputenc}
\usepackage{graphicx}
\usepackage{amssymb}
\usepackage{amsmath}
\usepackage{float}
\usepackage{dsfont}
\usepackage{overpic}

\RequirePackage{etoolbox}
\RequirePackage{keyval}
\RequirePackage{kvoptions}
\RequirePackage{ifthen}
\RequirePackage{url}
\RequirePackage{xstring}

\newcommand{\tr}[1]{\mathrm{tr}(#1)}

\begin{document}

\title{Device-Independent Bounds on Detection Efficiency}
\author{Jochen Szangolies}
\email{jochen.szangolies@hhu.de}
\author{Hermann Kampermann}\author{Dagmar Bru\ss}
\affiliation{Institut f\"ur Theoretische Physik III, Heinrich-Heine-Universit\"at D\"usseldorf, D-40225 D\"usseldorf, Germany}

\begin{abstract}
In many quantum information applications, a minimum detection efficiency must be exceeded to ensure success. Protocols depending on the violation of a Bell inequality, for instance, may be subject to the so-called detection loophole: imperfect detectors may yield spurious violations, which consequently cannot be used to ensure, say, quantum cryptographic security. Hence, we investigate the possibility of giving lower bounds on detector efficiency even if an adversary has full control over both the source and the detectors. To this end, we present a technique to systematically derive Bell inequalities free from the detection loophole using only the observed measurements statistics. The violation of these inequalities certifies that the detectors used exceed a certain minimal efficiency.

\end{abstract}

\maketitle

Quantum theory allows to perform certain tasks that are infeasible in the classical realm. Often, this quantum advantage is derived from violating so-called Bell inequalities, which are constraints on the correlations achievable in {\it local realistic} theories \cite{bell1964}. Bell inequality violation has in recent years been shown to furnish a resource, enabling the performance of tasks such as unconditionally secure quantum cryptography \cite{ekert1991}, exceeding classical performance in communication complexity tasks \cite{brukner2004}, and generating certifiably random numbers \cite{pironio2010}.

For such applications, it is necessary to exclude spurious violations of Bell inequalities, which are not due to the failure of local realism, but instead stem from experimental imperfections or unjustified additional assumptions. Such violations may occur, for example, due to improper causal separation of the apparatuses (the locality loophole \cite{bell1980}) or too low detector efficiency (the detection/fair sampling loophole \cite{pearle1970,clauser1974}). It has only recently become possible to simultaneously close these loopholes in actual experiments (\cite{hensen2015, shalm2015, giustina2015}).

In this paper, we suggest a method to bound detection efficiencies even in the presence of adversarial influences. To this end, we exhibit a new way to construct Bell inequalities based only on the observed measurement outcomes. 

Finding such Bell inequalities is interesting in itself, since by construction, we know that this violation cannot be due to sampling effects. Thus, where a setup using a pre-chosen Bell inequality may fail to produce a violation, e.g. due to noise issues, using our method, a (violated) Bell inequality will be found whenever the data is not compatible with a classical model.

As we will show, the violation of these Bell inequalities can be used to establish lower bounds on detector efficiencies even in the fully device-independent scenario. This is in contrast to the situation in classical physics, where variations in source rate or pre-programmed pseudo-detections always allow the `faking' of detector efficiencies.

Utilizing Bell inequality violations as a means to `self-test' experimental setups has been proposed before. Previous applications include verification of states and measurements \cite{may2003,yang2013,mckague2012}, the self-testing of quantum circuits \cite{mag2006}, and the certification of random numbers \cite{pironio2010}. Here, we propose a new self-testing task.

% Similarly, Branciard \cite{bra2011} considered the problem of finding Bell inequalities violated by detectors of a given detection efficiency. He defines a \emph{postselected polytope} containing correlations compatible with a local realistic model, given a certain efficiency of the detectors. Our approach, on the other hand, identifies Bell inequalities that do not suffer from the fair-sampling problem, which still bound the classical, local polytope (which is a proper subset of Branciard's postselected polytope). Furthermore, we exhibit a method to find a lower bound on detection efficiency even in a completely uncharacterised setting, by means of the Navascu{\'e}s-Pironio-Ac{\'\i}n (NPA) hierarchy \cite{NPA2007, NPA2008}.

{\it Bell inequalities and correlation polytopes.} Bell inequalities can be considered to stem from the insolubility of the so-called {\it marginal problem} in quantum settings \cite{fine1982}: in general, for a set of observables $\{A_i\}$, there exists no joint probability distribution $P(\{A_i\})$ such that its marginals recover the probability distributions of jointly measurable subsets of observables. The set of all probability distributions for which the marginal problem is solvable is a convex polytope \cite{pitowsky1989}; thus, Bell inequalities can be viewed as hyperplanes bounding this polytope. In this Letter, we will consider uncharacterised detectors. On this approach, the set of distributions for which nonclassicality cannot be certified is a convex cone, rather than, e.g., a polytope that is a superset of the polytope of classical correlations, as in the approach of Ref.~\cite{bra2011}.

We assume an experimental setup consisting of a source $\mathcal{S}$ and two detectors $\mathcal{A}$ and $\mathcal{B}$, belonging to Alice and Bob, respectively. The detectors are causally separated, and likewise, the source cannot be influenced by the detectors. 

We will in the following consider only dichotomic observables, and it suffices to restrict our attention to the $+1$-outcomes. We will write $p(A_i^+)$ ($p(B_j^+)$) for the probability that the $i$th observable of Alice (the $j$th observable of Bob) yields the value $+1$, and $p(A_i^+B_j^+)$ for the joint probability that both yield $+1$, where $i=1,\dots,n$ and $j=1,\ldots,m$ for arbitrary $n$ and $m$. All of these probabilities will be collected into probability vectors.

The polytope of classical correlations can be characterized by its extremal points $\mathbf{v}_k$, $k=1,\ldots,2^{n+m}$, i.e. those probability vectors whose entries are either $1$ or $0$. To derive these vertices, it suffices to note that $p(A_i^+B_j^+)=1$ if and only if $p(A_i^+)=1$ and $p(B_j^+)=1$. Every classically allowed probability distribution can then be written as a convex combination of these vertices, that is\\

\begin{equation}\label{pvec}
 \mathbf{P}_{\mathrm{class}}=(\mathbf{P}_A,\mathbf{P}_B,\mathbf{P}_{AB})^T=\sum_{k=1}^{2^{n+m}} \lambda_k \mathbf{v}_k,
\end{equation}
where $\lambda_k \geq 0$, $\sum_k \lambda_k=1$, and $\mathbf{P}_A=(p(A_1^+),p(A_2^+),\ldots,p(A_n^+))^T$ denotes the vector of probabilities for Alice's observables to yield $+1$, and analogously for Bob's probability vector $\mathbf{P}_B$ and the joint probability vector $\mathbf{P}_{AB}$.

Conversely, any probability distribution that does not admit such a decomposition violates at least one Bell inequality. Using the convex decomposition into vertices of the polytope, the question of classicality of a probability distribution can then be answered using a linear program.

{\it Bell inequalities without fair sampling assumption.} Consider the following scenario: you are at the used-detector merchant of your choice, and want to pick a detector meeting your requirements regarding detection efficiency. However, all of the equipment is under control of the vendor. Since the vendor has a vested interest in selling you his equipment (and all sales are final), you thus need to implement a protocol that allows you to assess the detector's quality in a way secure against tampering by the vendor.

Choosing some Bell inequality in advance is likely to be inefficient, as it will typically not be violated, even if the prepared state is entangled. Hence, we propose to directly construct Bell inequalities from any observed probability distribution by means of a linear program. %To do so, we find a linear program that determines, for the experimentally observed probabilities, whether they can be explained by a local realistic model. If that is not the case, a Bell inequality violated by this probability distribution is produced. 

It has been shown previously that random local measurements can be used to generate Bell inequality violations, thus obviating e.g. the need for a shared reference frame between distant experimenters \cite{liang2010,wallman2011,wallman2012,shadbolt2012}.

In any real experiment non-detections are present, such that sampling effects may induce Bell inequality violations not present if the whole ensemble were taken into account (detection loophole \cite{pearle1970}). 

Fortunately, the polytope method can be adapted for this case. We simply need to reformulate everything in terms of the the actually observed $+1$ outcomes for each observable \cite{eberhard1993}. Consider the Bell inequality
\begin{equation}\label{BI}
 \sum_{i,j=1}^{n,m}h_{A_iB_j}p(A_i^+B_j^+)+\sum_{i=1}^{n} h_{A_i}p(A_i^+) + \sum_{j=1}^{m} h_{B_j}p(B_j^+)\leq c,
\end{equation}
where $c$ is the classical bound, and $h_{A_i}$, $h_{B_j}$ and $h_{A_iB_j}$ are coefficients defining the Bell inequality. We can, for a large enough sample size $N$, replace the probabilities with the relative frequencies, e.g. $p(A_i^+)=N_{A_i}^+/N$, where $N_{A_i}^+$ is the number of occurrences of the $+1$-outcome upon measuring $A_i$ \footnote{Note that some care must be taken here: to avoid double counting, only $+1$-outcomes within a single experimental context must be considered.}. This yields the Bell inequality
\begin{equation}
  \sum_{i,j=1}^{n,m}h_{A_iB_j}N_{A_iB_j}^{++} + \sum_{i=1}^{n} h_{A_i}N_{A_i}^+ + \sum_{j=1}^{m} h_{B_j}N_{B_j}^+\leq Nc,
\end{equation}
where we have already multiplied by the total number of events $N$. %In a scenario with untrusted devices, this number is unknown; however, the observed counts $N_{A_i}^+$ are accessible. 
%We will assume, in the following, that the number $N$ is unknown: while we could simply define the instances where an experiment is carried out, and assign the value $-1$ whenever no detection is made, this will tend to drive down, and eventually, overwhelm any possible violation. 
Leaving $N$ open here allows us to solely use the observed counts $N_{A_i}^+$. For any Bell inequality with $c=0$, thus, the unknown $N$ drops out, and we are left with an inequality containing only directly observable quantities \footnote{As presented, this approach is strictly valid only for the case of an equal number of measurements in each direction; however, since this number is due to the choice of the experimenter, this does not pose a restriction. Furthermore, it is possible to adapt the approach to more general cases \cite{kof2016}.}. To work with probabilities again, we may divide by some base rate $N_{\mathrm{obs}}$---e.g. the total number of detections. This yields then our observed probabilities, given by the vector $\mathbf{P}_{\mathrm{obs}}$.

Our task now is to decide whether this $\mathbf{P}_{\mathrm{obs}}$ is a classical probability distribution.%; that is, whether there exists a Bell inequality of the form of Eq.~(\ref{BI}) with $c=0$ such that $\mathbf{P}_{\mathrm{obs}}$ violates this inequality.
This question can be formulated as a linear separation problem: find a hyperplane separating all the $\mathbf{v}_k$ and the point given by the observed probability distribution $\mathbf{P}_{\mathrm{obs}}$. A general hyperplane in $d$ dimensions containing the origin is given by 
\begin{equation}
 \sum_{l=1}^dh_lx_l=0,
\end{equation}
where the $x_l$ are Cartesian coordinates and the $h_l$ are elements of the hyperplane's normal vector $\mathbf{h}$. Hence, the problem of finding such a hyperplane translates to:

\begin{center}
\begin{tabular}{r l}\label{linp}
 find:       & $\mathbf{h}\in\mathbb{R}^d,\,d=n+m+nm$\\
 subject to: & $\mathbf{h}^T\mathbf{v}_k\leq0\,\forall\,k=1,\ldots,2^{n+m}$\\
             & $\mathbf{h}^T\mathbf{P}_{\mathrm{obs}}>0$.\\
\end{tabular}
\end{center}

\begin{figure}[h]
 \centering
%  \begin{overpic}[width=0.7\columnwidth%,grid,tics=10
%   ]{corrspace}
%   \put(42,45){classical}
%   \put(41,27){quantum}
%   \put(36,18){nonsignalling}
%   \put(49,72){$\mathbf{v}_1=0$}
%   \put(16,28){$\mathbf{v_2}$}
%   \put(81,28){$\mathbf{v_3}$}
%   \put(80,60){$\mathbf{h}$}
%  \end{overpic}
 \includegraphics[width=0.8\columnwidth]{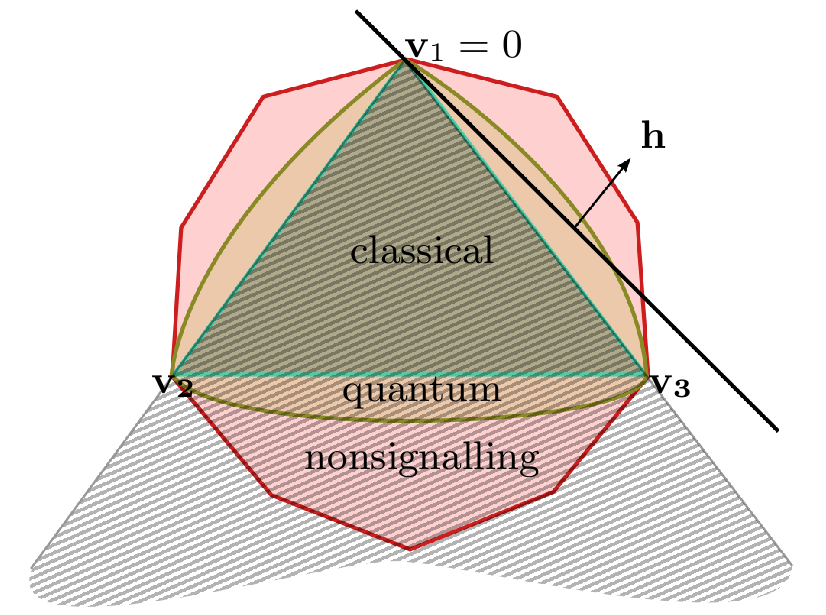}
 \caption{The sets of classical, quantum, and nonsignalling correlations, a Bell inequality defined by its normal vector $\mathbf{h}$, and the cone of probability distributions where we cannot exclude the existence of a classical model (hatched area).}
 \label{corrspace}
\end{figure}

The hyperplane then defines the Bell inequality 
\begin{equation}\label{BIs}
 \sum_{l=1}^{n+m+nm}h_lp_l\leq 0,
\end{equation}
where $p_l$ are the elements of the observed probability vector $\mathbf{P}_{\mathrm{obs}}$, and the $h_l$ yield the coefficients of the Bell inequality as in Eq.~(\ref{BI}). If the linear program has a solution, then the observed probability distribution violates this Bell inequality, and is free from the fair sampling loophole. To find the Bell inequality with the maximum quantum value, we optimize the quantum value $Q=\mathbf{h}^T\mathbf{P}_{\mathrm{obs}}$, with the additional constraint of $-1\leq h_l\leq 1$ to keep the problem bounded, which merely introduces an arbitrary scale. The geometry of the situation is schematically shown in Fig.~\ref{corrspace}. The figure includes the set of general nonsignalling distributions, %i.e. the set of all probability distributions consistent with the impossibility of sending a signal between $A$ and $B$, 
which is a superset of the set of quantum correlations \cite{popescu1994}.

As an example, the well-known CH-inequality \cite{clauser1974} is defined by the hyperplane with normal vector $\mathbf{h}=(-1,0,-1,0,1,1,1,-1)^T$.%, yielding
% \begin{align}
%  \nonumber\sum_{l=1}^8 h_lp_l=&-p(A_1^+) - p(B_1)^+ + p(A_1^+B_1^+) + p(A_1^+B_2^+) \\&+ p(A_2^+B_1^+) - p(A_2^+B_2^+)\leq 0.
% \end{align}

The method as outlined so far already has several interesting applications. First, it can be considered a further development of the protocols in Refs.~\cite{liang2010,wallman2011,wallman2012,shadbolt2012}, %obviating not just the need for a global shared reference frame, but 
achieving Bell inequality violation without any characterization of the devices involved, thus making it device-independent. In Ref.~\cite{wallman2012}, it is shown that three measurements per party along orthogonal axes of the local coordinate system always suffice to yield a Bell inequality violation if both parties share a maximally entangled two-qubit state. 

To gauge the efficiency of our method for this application, we performed a numerical simulation of $5\cdot 10^5$ instances of attempting to generate a Bell inequality violation using up to $n=m=6$ randomly chosen measurements per party on maximally entangled two-qubit states. The simulation was performed using the MATLAB-toolboxes YALMIP \cite{lofberg2004} and SDPT3 \cite{toh1999}.

Despite the lack of characterization of the detectors in our case, more than half of all instances were successful using only $n=m=3$ measurements, while six measurements suffice in more than $99\%$ of all cases. Hence, despite needing fewer assumptions, our protocol's efficiency remains comparable to the one in Ref.~\cite{wallman2012}.

As a second application, our method represents a device-independent entanglement detection protocol for unknown states. Thus, it is a natural further development of the method presented in Ref.~\cite{szangolies2015}, removing the characterization of the detectors necessary therein.

For a further application, note that in device-independent quantum key distribution (DIQKD), the secret key rate $R$ is connected to the quantum value $Q>0$ of the Bell inequality used for security \cite{mas2011}:
\begin{equation}\label{ref}
 R\geq -\mathrm{log}_2f(Q)-H(a|b),
\end{equation}
where $f(Q)$ is a function depending on the Bell inequality used, and $H(a|b)$ is the conditional Shannon entropy of Alice's outcomes $a$ and Bob's outcomes $b$. 

Our method now suggests a DIQKD protocol in which the Bell inequality is not agreed upon beforehand, but rather, is constructed such that, given the observed probability distribution of local measurement outcomes, the quantity $R$ is maximized. This ensures both that a Bell inequality is chosen that leads to the best key rate given the actually performed measurements (which may differ from the measurements Alice and Bob set out to perform, either due to noise or the actions of an eavesdropper), and guarantees the closing of the fair-sampling loophole. 

We now turn towards the novel task of generating bounds on the efficiency of detectors in an adversarial scenario.

{\it Bounding detector efficiencies.} Let us now ask whether a violation of the inequalities we have deduced is still observable with some given limited detection efficiency. Here, by detection efficiency we mean the probability $\eta$ that a detector clicks on the arrival of a particle.

Including detector efficiencies, e.g. the probability for joint $+1$-outcomes reads $p(A_i^+B_j^+)=\eta_A\eta_B\tr{\rho_{AB}\Pi_{A_i}^+\otimes\Pi_{B_j}^+}$, where $\eta_A$ ($\eta_B$) is the probability that detector $\mathcal{A}$ ($\mathcal{B}$) fires, and $\Pi_{A_i}^+$ ($\Pi_{B_j}^+$) is the projector on the $+1$-eigenspace of $A_i$ ($B_j$). Thus, we find threshold detection efficiencies for each inequality: our inequalities~(\ref{BIs}) now read
\begin{align}
 \nonumber\eta_A\eta_B&\sum_{i,j=1}^{n+m}h_{A_iB_j}p(A_i^+B_j^+)\\+\eta_A&\sum_{i=1}^{n} h_{A_i}p(A_i^+) + \eta_B\sum_{j=1}^{m} h_{B_j}p(B_j^+)\leq0.
\end{align}
Assuming that $\eta_A=\eta_B\equiv\eta$, the critical detection efficiency is given by
\begin{equation}\label{crit}
 \eta_{\mathrm{crit}}=-\frac{\sum_i h_{A_i}p(A_i^+) + \sum_jh_{B_j}p(B_j^+)}{\sum_{ij}h_{A_iB_j}p(A_i^+B_j^+)}.
\end{equation}

Note that due to the fact that $\eta_{\mathrm{crit}}$ is a nonlinear function of the probabilities, we cannot use a straightforward SDP-approach to find the optimal value. One way around this is to implement the optimization by means of an iteration: set a fixed value for $\eta_{\mathrm{crit}}$, then check if we can still generate a Bell inequality violation, by optimizing over probability distributions possessing a quantum model. If this is the case, $\eta_{\mathrm{crit}}$ is decreased; otherwise, it is increased, until a value is found such that the Bell inequality just fails to be violated, in order to obtain a true lower bound. This procedure is still effectively implementable on a standard desktop computer up to at least $m=n=6$ local measurements.

We have here made an assumption that the efficiency of the detector does not depend on precisely which observable is being measured. This is justified for instance in the case where the detector is a simple photon-counter, and different observables are realized via different positioning of the detector in an optical experiment, or different optical elements. 

In order to obtain the minimum detection efficiency necessary to violate a given Bell inequality, we have to determine the probability distribution $\mathbf{P}_\mathrm{opt}$ such that $\eta_\mathrm{crit}$ is minimal. In general, optimization over the full set of quantum correlations is infeasible. However, to obtain a lower bound, we can utilize the Navascu{\'e}s-Pironio-Ac{\'\i}n (NPA) hierarchy \cite{NPA2007, NPA2008}, which yields a nested set of semidefinite criteria for a given probability distribution to have a quantum model. If on the $k$-th level of the hierarchy, a certificate obeying certain conditions exists, then that distribution may admit a quantum model; if such a certificate does not exist, then the probability distribution cannot originate from a quantum experiment. 

Thus, each further level excludes more probability distributions, and hence, yields a better lower bound for the critical detection efficiency, reproducing the exact quantum bound in the infinite limit. In order to implement the NPA hierarchy, we used the freely available MATLAB-toolbox QETLAB \cite{qetlab}.

In practice, often using few levels suffices to obtain an accurate bound on $\eta_{\mathrm{crit}}$; for the CH-inequality, e.g., already the nonsignalling correlations (corresponding to the `$0$'th-level) yield a bound of $\eta_{\mathrm{crit}}=\frac{2}{3}$. As shown in Refs.~\cite{eberhard1993,larsson2001}, this is indeed the optimal bound. 

An advantage of this method is that it yields lower bounds on the critical detection efficiency for arbitrary Bell inequalities. For example, the Bell inequality $\sum_lh_lp_l\leq0$ with $n=6$ and $m=5$ measurements found using our method by performing random unit-efficiency measurements on a maximally entangled two-qubit state, defined by the coefficients
\begin{align}
            \label{I6522a}           
            (h_{A_i})^T&=(-4,-6,-6,-4,-6,0)\\
            \label{I6522b}           
            (h_{B_j})^T&=(-2,-6,-4,-6,-6)\\
            \label{I6522c}           
            (h_{A_iB_j})&=\begin{pmatrix}      
                           6 &  0 &  2 &  2 & -2 \\
                          -6 &  6 &  6 &  2 &  4 \\ 
                           0 &  3 & -2 &  5 &  5 \\
                           0 & -3 & -2 &  6 &  6 \\
                           6 &  6 &  0 & -6 &  6 \\
                          -2 &  0 &  4 &  4 & -6
                        \end{pmatrix}, 
\end{align}

by using the NPA-hierarchy up to the $2$nd level, leads to a lower bound of $\eta_{\mathrm{crit}}>0.86$.

Dropping the assumption of equal detection efficiencies and rather assuming the worst case, namely one perfect detector, non-trivial lower bounds are still possible: the inequality defined by the coefficients in Eqs.~\ref{I6522a}--\ref{I6522c} yields a bound of $\eta_{A,\mathrm{crit}}>0.751$. This can be further improved by using, instead of the classical bound $0$, the observed quantum value $Q$, yielding  
\begin{equation}
 \eta_{A,\mathrm{crit}}=\frac{Q-\sum_jh_{B_j}p(B_j^+)}{\sum_{ij}h_{A_iB_j}p(A_i^+B_j^+)+\sum_ih_{A_i}p(A_i^+)}.
\end{equation}
In our simulation using random measurements, a value of $Q=1.971$ was produced (where an upper bound to the maximal value, obtained at the $2$nd level of the NPA-hierarchy, is $Q_2=3.6791$), which yields $\eta_{A,\mathrm{crit}}>0.886$.

We tested our method by implementing $10^3$ simulations of the setting with $m=n=2$ local observables, and detectors operating at efficiency $\eta=0.9$. Repeating the simulated experiment with random measurement directions on a maximally entangled state until a Bell inequality violation was obtained, we found that, taking account of the quantum violation $Q$ in each case, at the second level of the NPA-hierarchy, we could reconstruct an average detection efficiency of $\eta\geq 0.785\pm0.003$, where the uncertainty is due to the finite sample size. 

The method as presented so far assumes a quantum source for the observed data. However, it is simple to relax this assumption, instead e.g. merely requiring that the probabilities be compatible with the no-signalling constraint. Doing so leads to a lower minimal detection efficiency; in the simulation described in the previous paragraph, we are then able to reconstruct a detection efficiency of $\eta\geq 0.683\pm0.001$.

\begin{figure}[h]
\centering
% \begin{overpic}[width=\columnwidth%,grid,tics=10
% ]{Qdep3}
%  \put(45,2){$\eta_{\mathrm{known}}$}
%  \put(5,32){\rotatebox{90}{$\eta_{A,\mathrm{crit}}$}}
% \end{overpic}
\includegraphics[width=\columnwidth]{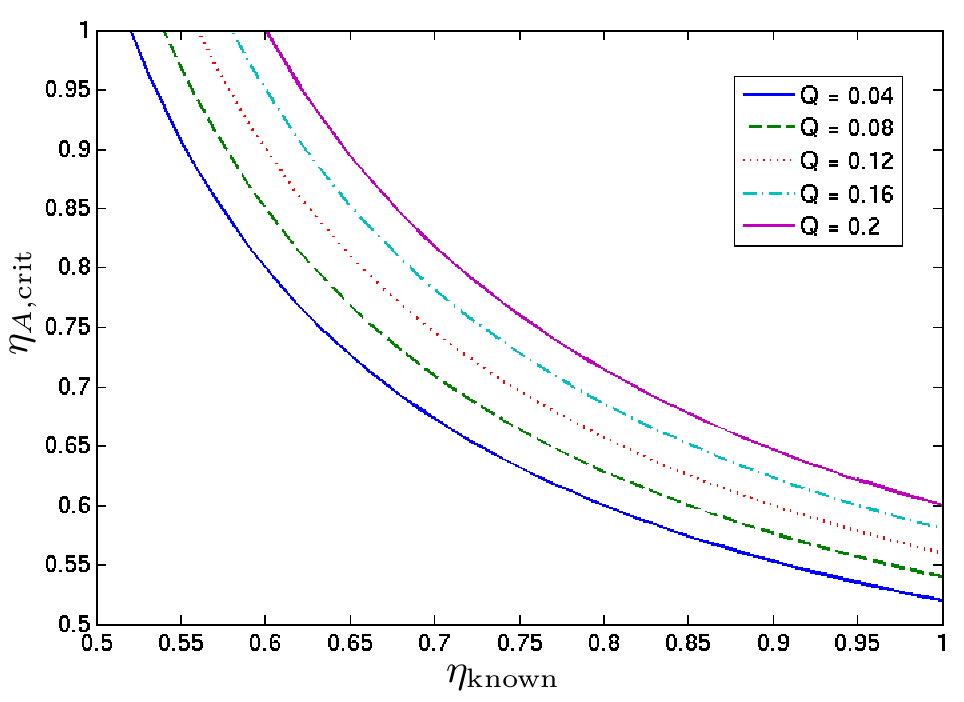}

\caption{Certified lower bound on the detection efficiency of an unknown detector versus the efficiency of the known detector using the CH-inequality for the indicated quantum values $Q$.}
\label{effQ}
\end{figure}

Another approach is to assume the availability of one detector with a known upper bound $\eta_{\mathrm{known}}$ on its detection efficiency---say, you have brought your old detector, which you want to replace with a better one. As an example, in Fig.~\ref{effQ}, the lower bound on the detection efficiency of the unknown detector is plotted against the efficiency of the known detector for testing the CH-inequality with a quantum value in the range of $Q=\{0.04,0.08,0.12,0.16,0.2\}$.

{\it Conclusions.} We have demonstrated a method to systematically derive Bell inequalities immune to the fair sampling-loophole, based only on the experimental data. Our linear program checks %whether the observed statistics belong to the convex polytope of local realistic correlations, and thus, 
whether a Bell inequality can be constructed that is violated by these probabilities. 

This method has several interesting applications. It can be used to remove the assumptions on the characterization of detectors previously necessary to generate Bell inequality violations for parties that do not share a common reference frame; to detect the entanglement of unknown quantum states in a device-independent way; and to obtain bounds on secret key rates in DIQKD scenarios where both parties do not have to agree on a Bell inequality beforehand.

Furthermore, we discussed how this method can be used to derive bounds on the efficiency of detectors in an adversarial setting, a novel problem which does not have a classical solution. After constructing a (violated) Bell inequality, the critical efficiencies of the detectors necessary to violate the constructed Bell inequality may be computed, thus allowing to certify a lower bound on the detector's efficiency.

\bibliography{detbib}{}
\bibliographystyle{apsrev4-1}

\end{document}